# Numerical simulation of a rotating magnetic sail for space applications

Mingwei Xu[1], Ronghui Quan[1*], Yunjia Yao[1]
[1]Nanjing University of Aeronautics and Astronautics, Nanjing, Jiangsu, 210016
*Corresponding Author. Email: quanrh@nuaa.edu.cn

**Abstract**:
The Magnetic Sail is a space propulsion system that utilizes the interaction between solar wind particles and an artificial dipole magnetic field generated by a spacecraft's coil to produce thrust without the need for additional plasma or propellant. To reduce the size of the sail while improving the efficiency of capturing solar wind, a new type of rotating magnetic sail with an initial rotation speed is proposed. This study evaluates the thrust characteristics, attitude, and size design factors of a rotating magnetic sail using a 3-D single-component particle numerical simulation. The results show that an increase in rotational speed significantly increases the thrust of the rotating magnetic sail. The thrust is most significant when the magnetic moment of the sail is parallel to the direction of particle velocity. The study also found that the potential for the application of the rotating magnetic sail is greatest in orbits with high-density and low-speed space plasma environments. It suggests that a rotating magnetic sail with a magnetic moment (Mm) of $10^3$-$10^4$ $Am^2$ operating at an altitude of 400 km in Low Earth Orbit (LEO) can achieve a similar thrust level to that of a rotating magnetic sail operating at 1 AU (astronomical unit) of $10^7$-$10^8$ $Am^2$.

## 1. Introduction

Magnetic sail is a deep space propulsion system proposed by Zubrin and Andrews[1]. The artificial magnetic cavity (magnetosphere) formed by large-scale superconducting coils tens of kilometers in size is used to deflect the flow of solar particles. The momentum change of the solar wind is transmitted to the spacecraft to generate propulsion. The system does not require any propellant, so the ideal specific impulse of a magnetic sail is close to infinity. Additionally, the absence of propellant consumption also leads to a significantly longer lifetime compared to propellant-based propulsion systems.

Later, to reduce the structural size, Winglee et al. proposed the Mini-Magnetospheric Plasma Propulsion (M2P2) and Magnetic Plasma Sail (MPS) that cause magnetic layer expansion by plasma injection[2]. Slough proposed the plasma magnet drive system based on the rotating magnetic field generated by the plasma magnet[4]. The plasma magnet is composed of a pair of polyphase coils that generate a rotating magnetic field to drive the plasma necessary for propulsion, to expand and maintain the large magnetic structure[3]. Multiple research groups in Japan, as well as other research groups led by Funaki, have also carried out important work in magnetic sail propulsion systems [5]. Over time, the design of magnetic sail space propulsion devices has steadily evolved, and the first magnetic sail spacecraft model has been created [8]. Recently, G. Davidson and Vorobieff [9] proposed a magnetic sail concept that obtains propulsion force and charge from large-scale electromagnetic fields, such as the Magnetosphere, Heliosphere, or Interstellar Magnetic Field (ISMF).

The magnetic plasma propulsion system is smaller and lighter in structure, with higher engineering feasibility[2,3]. However, the injection of plasma makes it a propellant-based propulsion system, lacking the application advantage compared to traditional electric propulsion technology. In order to achieve the goals of reducing size, not consuming additional propellant, increasing the thrust-to-weight ratio, and reducing costs, this paper proposes the concept of rotating magnetic sail, as shown in Figure 1. By applying an initial spin angular velocity to the coil, the dipole magnetic field of the sail rotates around the magnetic moment axis. This method increases the momentum of the interaction between the solar wind and the sail, thereby increasing the resistance obtained by the magnetic sail. Unlike the magnetic field produced by a two pole stator winding[4], the magnetic dipole axis of the rotating magnetic sail remains fixed, and what changes is the velocity of the solar wind plasma relative to the magnetic sail propulsion. The advantage of this method is that it maintains the advantage of a non-propellant propulsion system, and at the same time, it increases the thrust-to-weight ratio of the pure magnetic sail while keeping the structure simple.

This study used a three-dimensional single-component particle simulation to describe the interaction process between a rotating magnetic sail and solar wind plasma, and analyzed the factors that affect the thrust characteristics of the rotating magnetic sail. These factors include solar wind ion velocity, sail rotation speed, pitch angle, and magnetic moment size. By studying these factors, the design of the rotating magnetic sail can be optimized to achieve higher thrust and better performance. In addition, this study also analyzed the application of the rotating magnetic sail in different interplanetary orbits and discusses how to reduce the size of the rotating magnetic sail while maintaining its thrust characteristics. Note that the thrust characteristics discussed in this paper include not only drag, but also lift and lateral force-induced steering angles and pitch moments. Lift and transverse force are evaluated as steering angle, while pitching moment. This is important for determining spacecraft attitude in mission design. The resulting thrust characteristic evaluation is helpful to further optimize the thrust prediction of the miniaturized magnetic sail propulsion system.

FIGURE 1: Concept of the rotating magnetic sail.

## 2. Numerical Model and Calculation Conditions

*2.1. Reference Coordinate System.* In order to explain the operating principle of the rotating magnetic sail, it is necessary to establish a rotating coordinate system. The interaction between solar wind particles and the dipole magnetic field is simulated in three-dimensional space. The coordinate system used in the calculations is shown in Figure 2. The z-axis of the stationary coordinate system O-XYZ is parallel to the direction of the incoming solar wind particles, which is also the direction of the generated propulsion force. The XY-plane is perpendicular to the flow direction. It is assumed that the origin of the rotating cylindrical coordinate system $O_R$-X'Y'Z' coincides with that of the stationary coordinate system, and there exists a constant angular velocity $\omega$ that rotates counterclockwise around the Y'-axis. The angle of attack $\theta$ is set as the angle between the Y'-axis and the Y-axis.

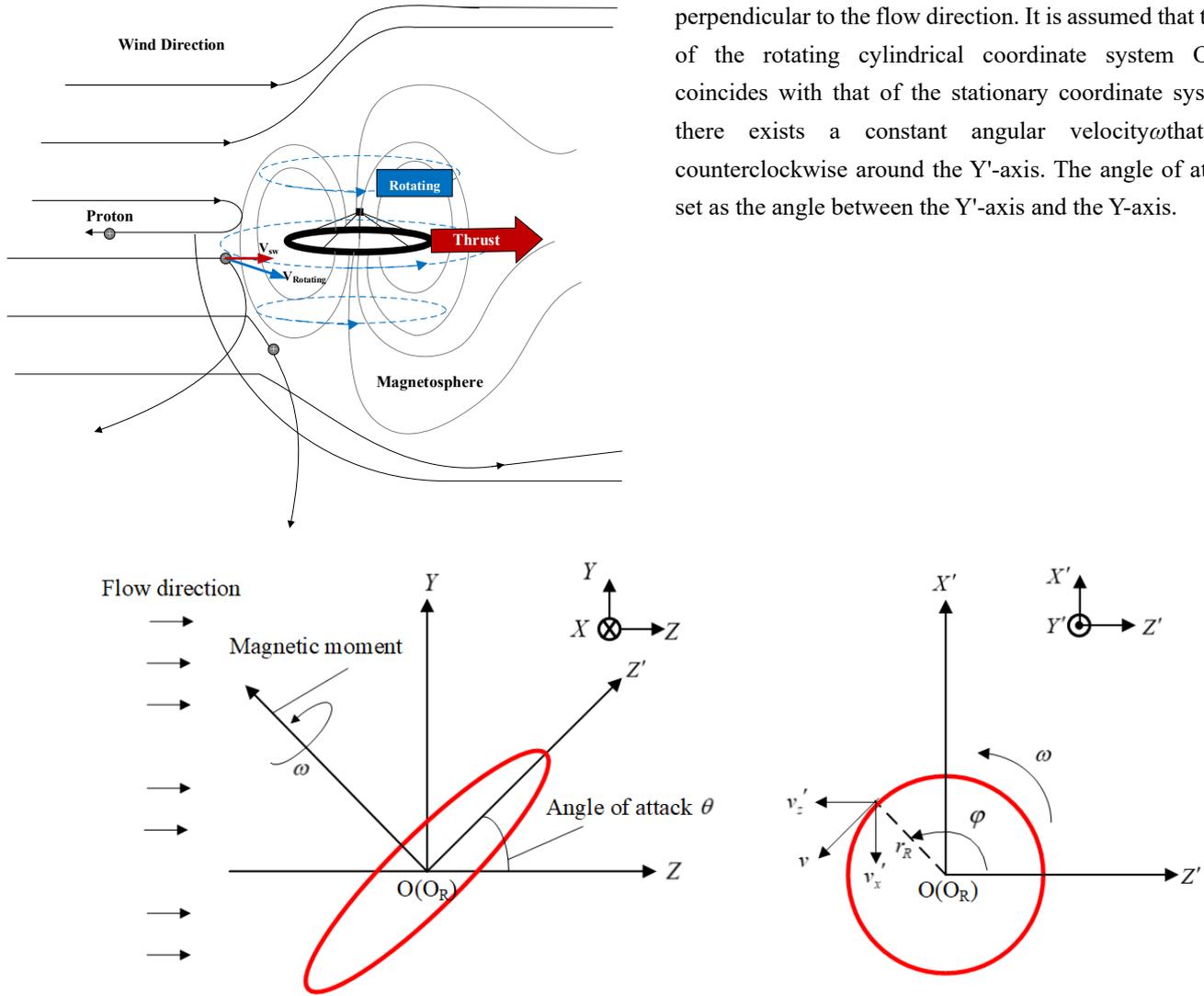

FIGURE 2: Rotating coordinate system and stationary coordinate system. The red loop represents the magnetic sail

Assuming that counterclockwise rotation is positive, the linear velocity in the $O_R$-X'Y'Z' coordinate system can be decomposed in the X'Z'-plane as follows:

$$\begin{cases} v_x' = -\omega r_R \sin\varphi \\ v_z' = -\omega r_R \cos\varphi \end{cases} \quad r_R = \sqrt{x'^2 + z'^2}, \varphi = \arctan\left(\frac{-z'}{x'}\right) \quad (1)$$

where $\omega$ is determined by the frequency $f$ of the drive motor ($\omega = 2\pi \cdot f$). The angle of attack, as shown in the right sub-figure of Figure 2, is defined as the angle between the Z' axis and the Z axis. The magnetic sail axis is perpendicular to the direction of solar wind ion flow when the angle of attack is 0 deg. When the Y' axis is not coincident with the Y axis, the rotational velocity coordinates are transformed to the O-XYZ coordinate system using the following :

$$\begin{cases} v_x = -\omega\sqrt{x^2 + (z\cos\theta - y\sin\theta)^2}\sin\left(\dfrac{y\sin\theta - z\cos\theta}{x}\right) \\ v_y = \omega\sqrt{x^2 + (z\cos\theta - y\sin\theta)^2}\cos\left(\dfrac{y\sin\theta - z\cos\theta}{x}\right)\sin\theta \\ v_z = -\omega\sqrt{x^2 + (z\cos\theta - y\sin\theta)^2}\cos\left(\dfrac{y\sin\theta - z\cos\theta}{x}\right)\cos\theta \end{cases}$$

(2)

As shown in Equation(3), **V** is the absolute velocity of the particle observed in O-XYZ, where $\mathbf{u_i}$ ($u_x$, $u_y$, $u_z$) is the velocity of the particle when there is no rotation.

$$\begin{cases} V_x = u_x + v_x \\ V_y = u_y + v_y \\ V_z = u_z + v_z \end{cases}$$

(2)

Note that here the velocities $\mathbf{v}(v_x, v_y, v_z)$ are superimposed at the moment of particle interaction with the magnetic field.

When evaluating the thrust characteristics of a magnetic sail, it is necessary to introduce important characteristic parameters: the representative length of magnetosphere $L$ and the Larmor radius $r_{Li}$. The magnetic moment of the sail and the plasma parameters of the solar wind determine the value of $L$, which can be determined through the balance between the magnetic and plasma pressures, as shown in Equations (4) to (5).

$$\frac{1}{2}nm_i u^2 = \frac{B_{mp}^2}{2\mu_0}\left(\frac{\mu_0 R_{coil} I_{coil}}{2\pi L^2}\right)^2$$

(3)

$$L = \left(\frac{\mu_0 M_m^2}{8\pi^2 nm_i u^2}\right)^{\frac{1}{6}}$$

(4)

Where $B_{mp}$ is the field strength at the magnetic field boundary, $n$ is the solar wind plasma density, $u$ is the solar wind flow velocity, $m_i$ is the ion mass, and **B** is the magnetic flux density vector, $\mu_0$ is the vacuum permeability, $M_m$ is the coil magnetic moment.

The ion Larmor radius $r_{Li}$ is the scale at which plasma particles are affected by the magnetic field, determined by Equation (6). It depends on the plasma density, velocity, and magnetic field strength. For typical solar wind parameters of $n = 5\times10^6$ m$^{-3}$ and $u = 5\times10^5$ m/s, and a magnetic flux density of approximately 36.2 nT at the magnetic field boundary $B_{mp}$, the ion Larmor radius at 1 AU can be calculated to be approximately 72 km.

$$r_{Li} = \frac{mu}{e \cdot 2B_{mp}}$$

(5)

$$\mathbf{B}_0 = \frac{\mu_0 I}{4\pi}\int_{L=2\pi r_R}\frac{dl_p \times (\mathbf{r}-\mathbf{r}_p)}{|\mathbf{r}-\mathbf{r}_p|^3}$$

(6)

$L$ represents the coil circuit and $I$ represents the coil current. Because of the symmetry of the calculation domain, when the

The simulation object of this study is the thrust characteristics of a magnetic sail with small-scale magnetic fields ranging from several hundred meters to several kilometers ($L\sim1000$m) [4]. In interstellar space, the average free path of ions ($\lambda = 1/n_i e \approx 1.25\times10^9$ km) is much longer than length of magnetosphere $L$[10]. Therefore, direct collisions between particles can be ignored. As $L$ is smaller than the ion Larmor radius, ions will enter the magnetic field region, and thus, it is necessary to consider the particle kinetics of ions and establish an ion motion model[8].

*2.2. Governing Equation.* Previous theoretical analysis of the thrust characteristics of the magnetic sail showed that, under the simulation condition of ignoring the time-varying electromagnetic field, the magnetosphere structure of the magnetic sail is mainly formed by the interaction between ions and the dipole magnetic field, and electrons hardly affect the structure of the magnetosphere[8]. This is because the induced current of ions and electrons is very weak and cannot produce an induced magnetic field equivalent to the dipole magnetic field of the coil. Ashida et al. obtained ion density distribution similar to self-consistent Full-PIC simulation through single-particle simulation tests[8]. The results show that the ion momentum change calculated by single-particle simulation is basically consistent with that calculated by Full-PIC simulation. Hence, we use a single-component plasma model for numerical testing. In this model, the electrostatic effect is ignored, and only one type of particle (ion) is explicitly treated. The presence of a background electron (massless fluid) ensures quasi-neutrality, and it is assumed that the charge separation is balanced by the background electron. Only the motion of ions under a steady-state magnetic field is considered. Note that because the electronic dynamics is ignored in the single particle simulation, the electric field produced by charge separation and the magnetopause current mainly determined by electrons are not taken into account, which means that the electromagnetic force ($\mathbf{J}_{\text{total}}\times\mathbf{B}$) acting on the magnetic sail will be less than that in the Full-PIC simulation[8,11].

Below, we will describe the basic equations and quasineutral model used in the simulation. Firstly, the magnetic field vector $\mathbf{B}_0$ of the coil dipole inside the simulation domain is calculated using the Biot-Savart law as follows, where $\mathbf{r}_p$ represents the position vector of the particle at a certain point inside the simulation domain[7]:

radius of the $r_R$ coil is in the XOZ plane and the particle potential is in the YOZ plane, $\mathbf{B_0}$ can be decomposed into:

$$B_x = \frac{\mu_0 I r_R}{4\pi} \int_0^{2\pi} \frac{-r\cos\vartheta \sin\varphi \, d\varphi}{\left(r_R^2 + r^2 + 2r_R r \sin\vartheta \cos\varphi\right)^{3/2}} = 0$$

$$B_y = \frac{\mu_0 I r_R}{4\pi} \int_0^{2\pi} \frac{(r_R + r\sin\vartheta \cos\varphi) \, d\varphi}{\left(r_R^2 + r^2 + 2r_R r \sin\vartheta \cos\varphi\right)^{3/2}} \quad (7)$$

$$B_z = \frac{\mu_0 I r_R}{4\pi} \int_0^{2\pi} \frac{-r\cos\vartheta \cos\varphi \, d\varphi}{\left(r_R^2 + r^2 + 2r_R r \sin\vartheta \cos\varphi\right)^{3/2}}$$

Where $\vartheta$ represents the angle between the position vector and the magnetic moment of the coil.

The Maxwell's equations are solved to obtain the self-consistent electromagnetic field, assuming the quasi-neutrality condition $n_i = n_e$, which means $\nabla \cdot \mathbf{J}_p = 0$ [13]. Due to the low frequency nature of solar wind plasma, the transverse displacement current is neglected using the Darwin approximation[6,8]:

$$\nabla \times \mathbf{B}_p = \mu_0 \mathbf{J}_p = e\mu_0 n_i (\mathbf{u}_i - \mathbf{u}_e) \quad (8)$$

$$en_e(\mathbf{E} + \mathbf{u}_e \times \mathbf{B}) + k_B T_e \nabla n_e = 0 \quad (9)$$

$$\mathbf{B} = \mathbf{B}_0 + \mathbf{B}_p \quad (10)$$

$$\frac{\partial \mathbf{B}}{\partial t} = -\nabla \times \mathbf{E} \quad (11)$$

$$\mathbf{E} = \left(\frac{1}{en_i \mu_0} \nabla \times \mathbf{B}_p - \mathbf{u}_i\right) \times \mathbf{B} - \frac{1}{en_i} k_B T_e \nabla n_i \quad (12)$$

Here, $\mathbf{B}_p$ represents the induced magnetic field of the plasma and $\mathbf{B}_0$ is the dipole magnetic field of the coil. The second term in Equation (10) represents the pressure of the electron fluid, taking into account the electron temperature $T_e$. Since the massless electrons have a small impact on the magnetic sail thrust, to simplify the model, we assume isotropic electron temperature in the magnetic field. Substituting equation (13) into (12), we obtain equation (14):

$$\frac{\partial \mathbf{B}}{\partial t} = -\nabla \times \left\{\left[\frac{1}{en_i \mu_0} \nabla \times (\mathbf{B} - \mathbf{B}_0) - \mathbf{u}_i\right] \times \mathbf{B} - \frac{1}{en_i} k_B T_e \nabla n_i\right\} \quad (13)$$

The motion equation for each ion is solved using the Buneman-Baris method as follows[14]:

$$m_s \frac{d(\mathbf{u}_i + \mathbf{v'})}{dt} = \mathbf{F}_s \quad (14)$$

$$\frac{d\mathbf{x}_i}{dt} = (\mathbf{u}_i + \mathbf{v'}) \quad (15)$$

$$\mathbf{F}_s = q_s \left((\mathbf{u}_i + \mathbf{v'}) \times \mathbf{B}\right) \quad (16)$$

Here, $\mathbf{u}_i$ is the velocity vector of the ion inflow, $\mathbf{v'}$ is the rotation velocity vector of the magnetic sail, and $\mathbf{F}_s$ is the Lorentz force acting on the particle. It can be seen that rotation changes the velocity of the ion relative to the magnetic field, and the particle experiences a force change. Since the Lorentz force corresponds to the scattering of solar wind particles, it characterizes the process of momentum transfer. Equation (17) is rewritten according to Equation (3) as follows:

$$\begin{cases} F_{s,x} = q\left(B_z(u_{i,y} + v_y) - B_y(u_{i,z} + v_z)\right) \\ F_{s,y} = q\left(B_z(u_{i,x} + v_x) + B_x(u_{i,z} + v_z)\right) \\ F_{s,z} = q\left(B_y(u_{i,x} + v_x) - B_x(u_{i,y} + v_y)\right) \end{cases} \quad (17)$$

Using the finite element method to solve Equations (15)-(17), the trajectory of each particle interacting with the magnetic field of the magnetic sail is obtained. Finally, the momentum of the incident ion pulse over the entire simulation domain is integrated to obtain the total force at different angles related to the velocity.

$$F_{sw} = \oiint -m_i n_i |\mathbf{u}_i| (\mathbf{V} - \mathbf{u}_i) \cdot d\mathbf{S} \quad (18)$$

$d\mathbf{S}$ represents an integral on any surface. The thrust in Z-direction, lift in Y-direction, and transverse force in X-direction are derived by decomposing the vector of the total force.

*2.3. Simulation Model and Initial Condition.* The input parameters for the 3-D simulation are presented in Table 1. A cylinder with a radius of 1.2 km and a height of 2 km is used as the computational domain. Typically, it is usually required that the grid spacing *dx* of the calculation region is not much larger than the plasma Debye length ( $dx/\lambda_D < 3$ )[15]. In this case, a grid spacing of *dx*=30 m is chosen. For low plasma density conditions, the time step dt used to be much smaller than the ion cyclotron frequency $\omega_{ci}$, dt=1×10$^{-6}$ s( $dt < 0.01 \times 0.01\omega_{ci}$ )[16].

The vector magnetic field is solved using finite difference method with zero-gradient boundary conditions at the boundaries. During the simulation, electron and ion temperatures are kept constant and the electrical resistivity is set to zero[10]. Solar wind ions are released from the inlet boundary along the positive Z-axis with typical solar wind speed and density distribution, and open boundary conditions are used. Particle trajectories are computed using the magnetic field as the initial condition[11]. The particle motion is computed using the generalized alpha method for time-stepping. At each time step, the forces acting on each particle at its current position are computed from the magnetic field results, and the particle positions are updated accordingly. This process is repeated until the simulation reaches the specified end time. The parameters for the solar wind inflow and the computational domain are presented in Table 1[18].

TABLE 1: Simulation parameters[18].

| Solar Wind Parameters | |
|---|---|
| Incoming velocity/km·s$^{-1}$ | 500 |
| Plasma flow | Proton(ion) |
| Density/m$^{-3}$ | 5×10$^6$ |
| Plasma temperature/K | 10$^5$ ($T_e=T_i$) |
| Ion mass/kg | 1.672×10$^{-27}$ |
| Ion Larmor radius/km | 72 |
| Debye length/m | 47 |
| Ion cyclotron frequency | $\omega_{ci}$=2.94×10$^3$ |
| Interplanetary magnetic field/T | 0 |
| Coil parameters | |
| Coil radius/m | 100 |
| Coil current/A | 10$^4$ |
| Attack angle θ/degree | 0,10,20,30,40,50,60,70,80,90 |
| Rotational speed/rpm | 0,100,200,300,400,500,600,700,800,900,1000 |
| Simulation parameters | |
| Grid/dx | 30 |
| Time step/dt | <0.00001×$\omega_{ci}$ |

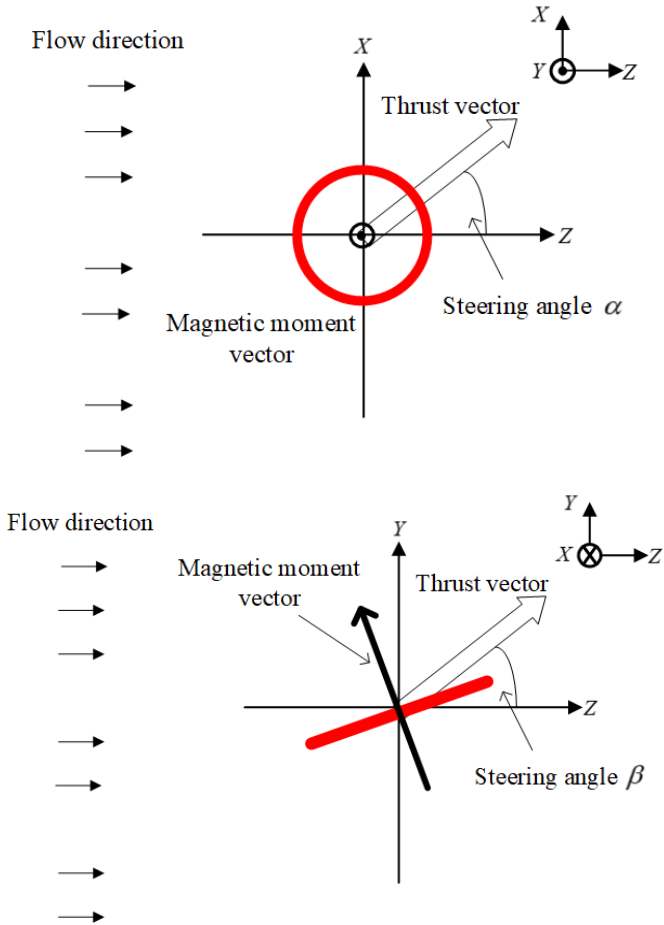

FIGURE 3: Definition of steering angles α and β.

Assuming the center of the coil is coincident with the origin of the coordinate axis, the simulation model evaluates the changes caused by rotation. When simulating, solar wind parameters (solar wind velocity $u_i$), magnetic sail coil design parameters (magnetic moment $M_m$, rotational speed $\omega_n$), and magnetic sail attitude parameters (angle of attack θ) are considered. Table 1 lists the relevant parameters for evaluating the thrust characteristics of the rotating magnetic sail. The rotational speed of the magnetic sail is set in the range of 0~1000 rpm (with a non-rotating case as a comparison). The range of attack angle for the magnetic sail is 0 deg to 90 deg. Figure 3 shows the turning angles α and β, represented by lift and transverse forces.

In order to verify the effectiveness of the simulation, first of all, in the case of no rotation, change the magnetic moment of the coil from 3×10$^6$Am$^2$ to 3×10$^{11}$Am$^2$ for simulation. Twelve groups of simulations are carried out with two stable states (θ = 0 deg and θ = 90 deg).

Figure 5 shows the results, with the vertical and horizontal axes representing the coil magnetic moment and the thrust of the magnetic sail, respectively. The straight line represents the least squares fit of full-PIC simulation results by Ashida [8], while the diamonds and triangles represent the single ion simulation results for attack angles of 0 deg and 90 deg, respectively. The thrust characteristics of the single-ion and full-PIC simulations are consistent when the magnetic moment is greater than or equal to 10$^8$ Am$^2$, but differ significantly when it is less than 10$^8$ Am$^2$. The thrust obtained from full-PIC simulation is two to three times that obtained from single ion simulation, as the latter neglects electron kinetic effects. According to the finite Larmor radius effect of electrons, the induced magnetic field and induced electric field increase, leading to an increase in drift velocity and induced current in non-uniform electromagnetic fields[10]. Therefore, if electron kinetic effects are considered, the thrust of a magnetic sail with small magnetosphere will become larger[8]. In the subsequent study, we select the case with better fitting performance ($M_m$ = 10$^8$Am$^2$) as the initial magnetic moment of the rotating magnetic sail.

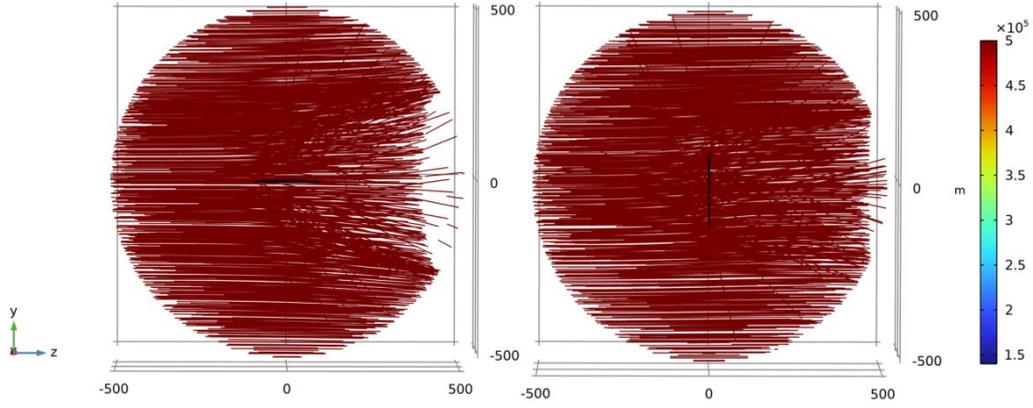

FIGURE 4: Ion trajectories without rotating magnetic sail along the YZ-plane for attack angle $\theta$ = 0 deg and $\theta$ = 90 deg.

The YZ-plane plots in Figure 4 show the ion trajectories at attack angle of 0 deg and 90 deg. The scale has been enlarged to clearly observe the wake regions where the ions interact with the magnetic field. Due to the weak induced electric and magnetic fields in the plasma, only ions close to the coil (origin) are affected by the strong dipole magnetic field. The particles are decelerated and deflected by the Lorentz force. As the Lorentz force is conservative, the magnitude of the particle velocity remains nearly constant. Both cases show wake structures that can be observed from inside the magnetic field and downstream of the magnetic sail, and the wakes are axisymmetric for both 0 deg and 90 deg. However, the wake area is larger at $\theta$ =0 deg than at $\theta$ =90 deg. At $\theta$ =90 deg, some particles will be completely reflected due to the magnetic mirror effect, which is a distinct feature that differs from the behavior observed at $\theta$ =0 deg.

$M_m$ and $\omega_n$. Under the conditions of magnetic moment $M_m$ = $10^7$ Am$^2$, $10^8$ Am$^2$, $10^9$ Am$^2$, a rotational angular velocity is applied to the magnetic sail for clockwise and counterclockwise rotations (designated as + and - respectively), ranging from 100 to 1000 rpm. Twelve simulations are conducted for the configurations of $\theta$ = 0 deg and $\theta$ = 90 deg. Figure 7 shows the ion trajectories at rotational speeds of $\omega_n$ = $\pm$100, 500, 1000 rpm for $M_m$ =$10^8$ Am$^2$, with each color legend representing the ion velocity for $\theta$ = 0 and $\theta$ = 90 deg. the wake region of rotating magnetic sail particles increases, the trajectory is more complex than that of non-rotating magnetic sails, and the particle velocity varies more widely. Counterclockwise rotation (+$\omega_n$) accelerates the particles near the coil, while clockwise rotation (-$\omega_n$) has the opposite effect, decelerating the particles. One characteristic related to rotational speed is that the wake appears to become wider as the rotation speed increases, indicating a greater number of deflected particles.

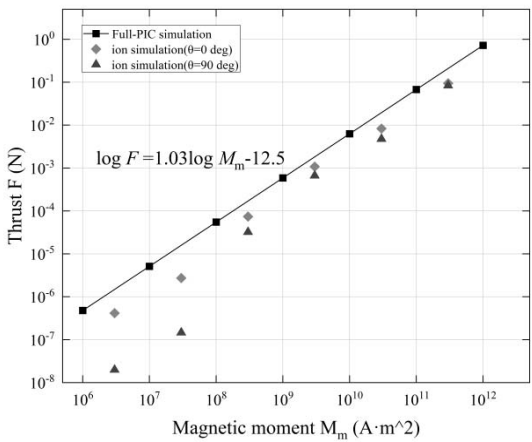

FIGURE 5: Thrust characteristics of pure magnetic sail of various magnetic moments obtained by 3-D simulation. ($\theta$ = 0 deg and 90 deg)[8].

## 3. Results and Discussion

*3.1. Thrust Characteristics Depending on Magnetic Moment*

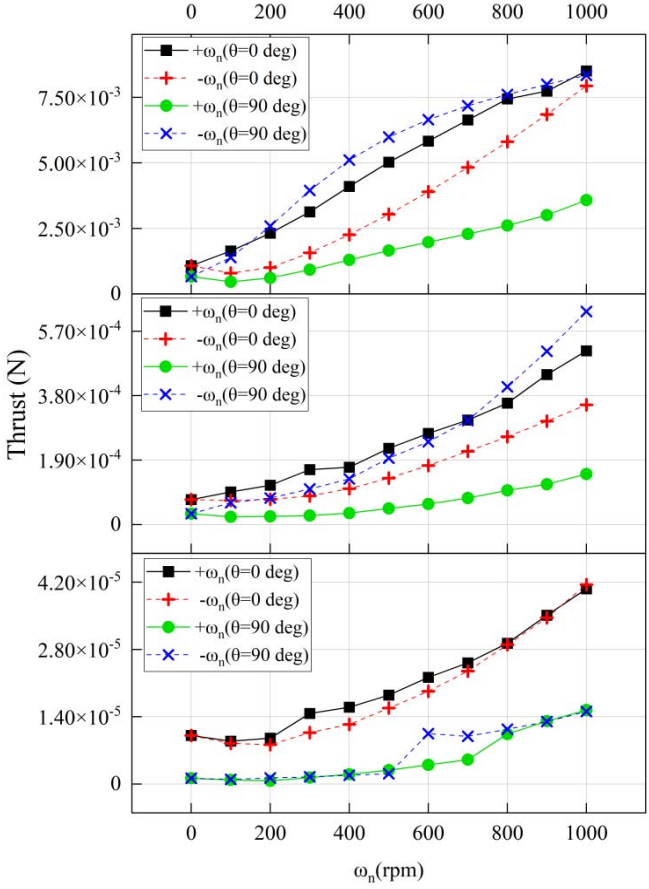

FIGURE 6: Thrust $F_z$ of rotating magnetic sail versus rotational speed $\omega_n$ for θ=0 deg and 90 deg: from top to bottom: $M_m = 10^7$ Am$^2$, $10^8$ Am$^2$, $10^9$ Am$^2$.

The Figure 6 shows the thrust $F_z$ for each group of magnetic moment parameters. The black square and red cross show the results for the $+\omega_n$ and $-\omega_n$ configurations, respectively, with an attack angle of 0 deg. The green circle and blue cross show the results for the $+\omega_n$ and $-\omega_n$ configurations, respectively, with an attack angle of 90 deg. From the figures, we can observe that:

(1) The thrust for both $\theta = 0$ deg and $\theta = 90$ deg configurations increases rapidly with increasing rotational speed.
(2) For the $\theta = 0$ deg configuration, the $+\omega_n$ rotation provides better gains, while for the $\theta = 90$ deg configuration, the $-\omega_n$ rotation provides better gains. The difference in thrust due to the rotation direction increases with increasing magnetic moment.

For $M_m=10^7$ Am$^2$, the thrust for the $\theta = 0$ deg configuration is significantly higher than that for the $\theta = 90$ deg configuration. The maximum thrust obtained for the $\theta = 0$ deg configuration is about 40.6 μN, which is 4.01 times the thrust for the no-rotation configuration (40.6μN/10.1μN). The maximum thrust obtained for the $\theta$= 90deg configuration is about 15.1μN, which is 12.83 times the thrust for the no-rotation configuration (15.1μN/1.2μN).

For $M_m=10^8$Am$^2$, the thrust for the $\theta = 90$ deg configuration is higher than that for the $\theta = 0$ deg configuration for rotational speeds greater than 700 rpm. The maximum thrust obtained for the $\theta=0$ deg configuration is about 0.512mN, which is 6.93 times the thrust for the no-rotation configuration (0.512mN/0.0739mN). The maximum thrust obtained for the $\theta = 90$ deg configuration is about 0.628mN, which is 19.38 times the thrust for the no-rotation configuration (0.638mN/0.0324mN). For $M_m=10^9$Am$^2$, the thrust for the $\theta=90$ deg configuration is higher than that for the $\theta = 0$ deg configuration for rotational speeds greater than 100rpm. The maximum thrust obtained for the $\theta=0$ deg configuration is about 8.51mN, which is 7.88 times the thrust for the no-rotation configuration (8.51mN/1.08mN). The maximum thrust obtained for the $\theta = 90$ deg configuration is about 8.34mN, which is 12.48 times the thrust for the no-rotation configuration (8.34mN/0.668mN).

Under the same design parameter settings, the thrust gain of a rotating magnetic sail is higher at θ= 0 deg than at $\theta = 90$ deg. This is because at $\theta = 90$ deg, the X and Y velocity components act on the particles initially moving along the Z direction via the Lorentz force, causing the particle trajectories to no longer move in a straight line along the Z-axis. As the rotational speed gradually increases, even in regions of weaker magnetic fields, the trajectories of the particles will change, causing the interaction time between the solar wind particles and the magnetic sail to be prolonged and the interaction region to become wider. In contrast, at $\theta=0$ deg configuration, only the x-direction velocity component is increased, resulting in a slightly smaller interaction region between the solar wind particles and the magnetic sail. Moreover, at low rotational speeds ($\omega_n=\pm100,200$ rpm) with smaller magnetic moments ($M_m=10^7$Am$^2$), the thrust change is not significant and may even decrease. However, as the magnetic moment increases, the thrust gain from rotation becomes more effective.

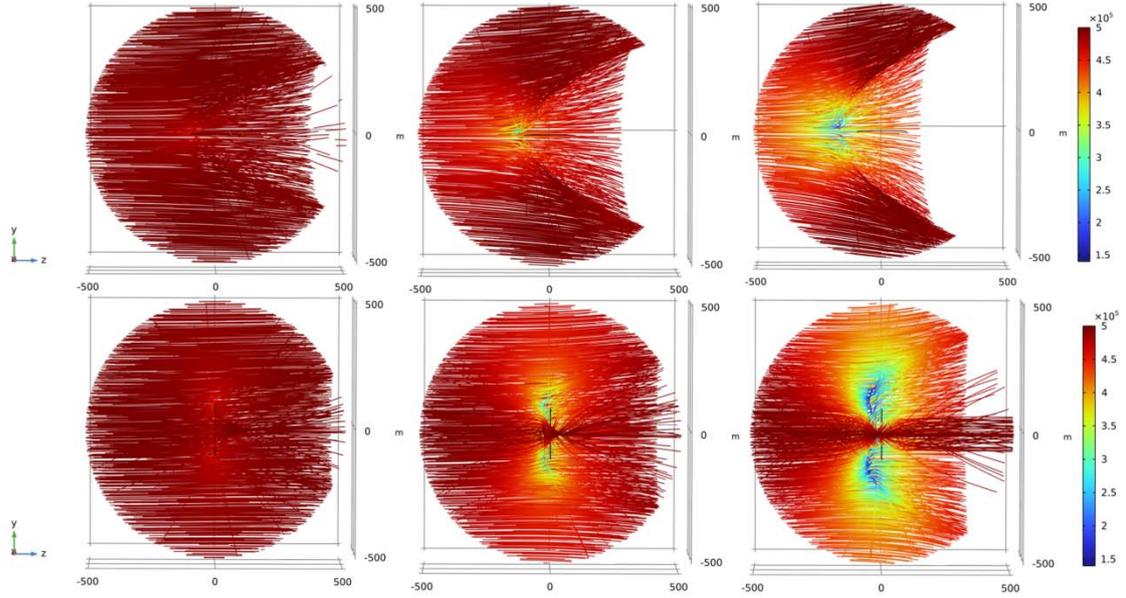

(a) Clockwise rotation speed, $+\omega_n$

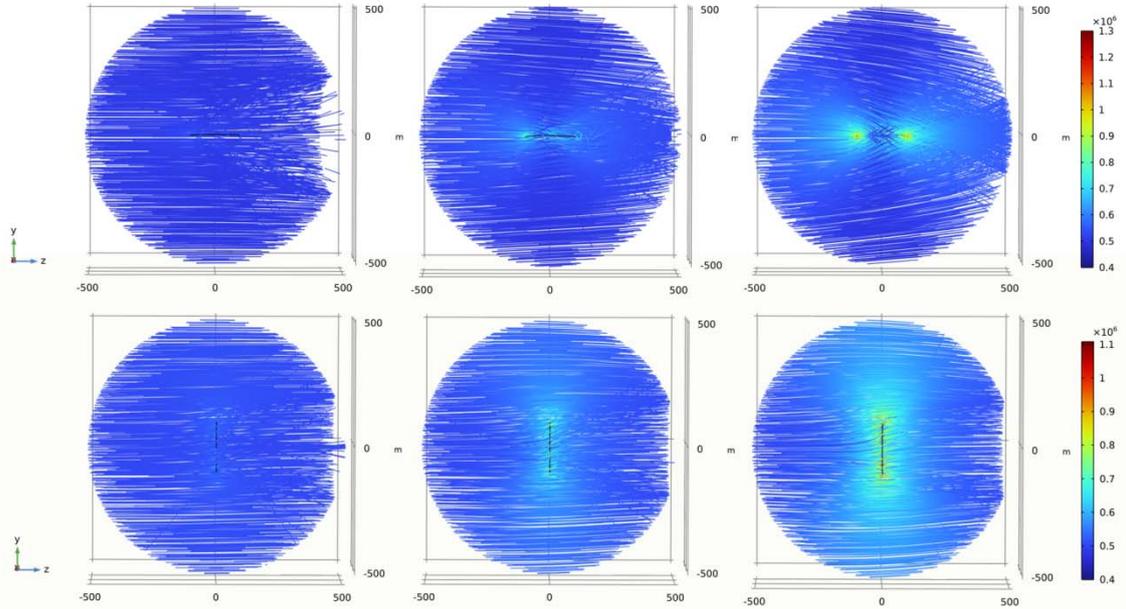

(b) Counterclockwise rotation speed, $-\omega_n$

FIGURE 7: Ion trajectories with rotating magnetic sail along YZ-plane for different rotation speed(from left to right: $\omega_n=\pm 100$ rpm, $\omega_n=\pm 500$ rpm, $\omega_n=\pm 1000$ rpm) and different angles(from top to bottom: $\theta = 0$ deg, $\theta = 90$ deg).

### 3.2. Thrust and Attitude Characteristics Depend on $\theta$ and $\omega_n$

*(1) Dependence of Thrust $F_z$ on on $\theta$ and $\omega_n$.* Three magnetic moments ($M_m = 10^7, 10^8, 10^9$ Am$^2$), six rotational speeds ($\omega_n = \pm 100, 500, 1000$ rpm), and a reference case of $\omega_n = 0$ are selected, along with nine angles of attack ($\theta = 0$ deg, 10 deg, 20 deg, 30 deg, 45 deg, 60 deg, 70 deg, 80 deg, 90 deg). The results for all other angles ($\theta = 90$ deg to $\theta = 360$ deg) can be derived from the results obtained for angles between $\theta = 0$ deg and $\theta = 90$ deg using the principle of symmetry. Figure 8 shows how the thrust $F_z$ obtained from the magnetic sail depends on the angle of attack $\theta$. In the case of no rotation, the thrust is almost monotonously decreasing with the increase of the angle in the range of magnetic moment checked, and the angular dependence of thrust is less than 3 times. In contrast, in the simulations with rotation, the dependence of the thrust on the angle is different for different magnetic moment conditions: for $M_m = 10^7$ Am$^2$, the obtained thrust gradually decreases as the angle changes from 0 deg to 90 deg. The thrust obtained for $-\omega_n$ is greater than that obtained for $+\omega_n$, and this difference increases as the rotational speed increases. For $M_m=10^8 \text{Am}^2$, under the $+\omega_n$ condition, the thrust decreases as the angle increases, while under the $-\omega_n$ condition, the thrust increases as the angle increases. There is a crossover point between the thrusts obtained for $-\omega_n$ and $+\omega_n$ at the same rotational speed, with $F_{+\omega n}>F_{-\omega n}$ for angles between 0 deg and 30 deg and $F_{-\omega n}>F_{+\omega n}$ for angles between 30 deg and 90 deg. The same pattern is observed for $M_m=10^9 \text{Am}^2$, and the crossover point shifts towards higher angles with increasing rotational speed, occurring at $\theta$ =42 deg ($\omega_n$=100 rpm), 45 deg ($\omega_n$=500 rpm), and 70 deg ($\omega_n$=1000 rpm). Therefore, when determining the direction of rotation of the magnetic sail, the angle of attack must be considered. At low magnetic moments or low rotational speeds, the difference in thrust caused by clockwise or counterclockwise rotation is not significant, but as the magnetic moment and rotational speed increase, rotation changes the angle-dependence of the thrust of the sail. By adjusting the direction and speed of rotation, the thrust magnitude and the sail attitude can be controlled.

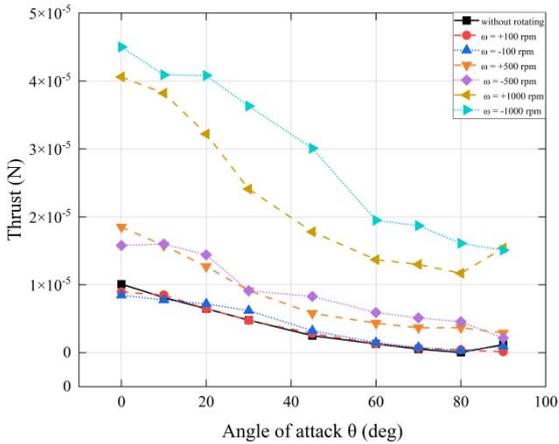

(a) $M_m = 10^7 \text{Am}^2$

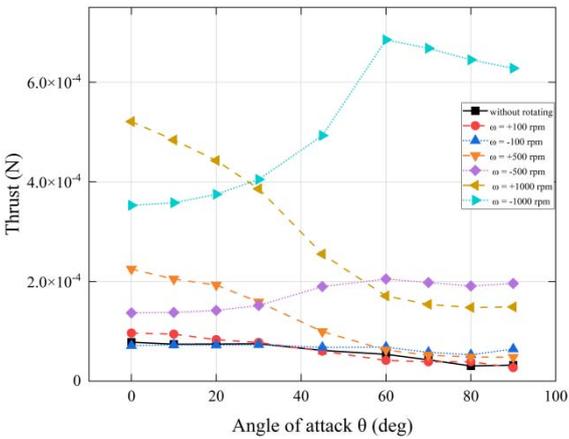

(b) $M_m = 10^8 \text{Am}^2$

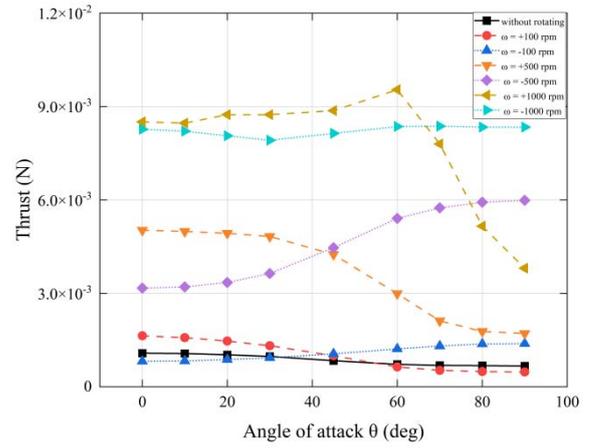

(c) $M_m = 10^9 \text{Am}^2$

FIGURE 8: Thrust of rotating magnetic sail versus the angle of attack $\theta$.

*(2) Steering Angles α and β.* Figures 9 and 10 show that when $M_m = 10^8 \text{Am}^2$, the steering angles $\alpha$ and $\beta$ are functions of the attack angle of the magnetic sail, which define how to control the attitude of the spacecraft and modify the orbit. The steering angle $\alpha$ is calculated from the arctangent of the ratio between the transverse force in the X direction and the thrust force in the Z -direction. The steering angle $\alpha$ is negative for clockwise rotation. From the plot, it can be seen that the rotation direction has little effect on the steering angle $\alpha$.

The extreme steering angle $\alpha$ is obtained at $\theta$ =0 deg, and it is -22 deg when there is no rotation. As the rotation speed increases, the extreme steering angle decreases. The dependence of the extreme steering angle $\alpha$ on the angle decreases, and at $\omega_n=\pm 1000$ rpm, the extreme steering angle $\alpha$ is -4 deg. The steering angle $\alpha$ represents the change in momentum in the XZ-plane, which means that as the rotation speed increases, the angle at which the spacecraft deviates from the -X direction decreases.

The steering angle $\beta$ is calculated from arctangent of the ratio between the transverse force in the Y-direction and the thrust force in the Z-direction. The steering angle $\beta$ is positive for counterclockwise rotation. From the plot, it can be seen that in the absence of rotation, the extreme steering angle $\beta$ is obtained at an angle of attack of 30 deg, which is 19 deg. In the case of $+\omega_n$, the extreme steering angle is obtained at an angle of attack of 30 deg, while in the case of $-\omega_n$, the maximum deflection angle is obtained at an angle of attack of 45 deg. Moreover, as the rotation speed increases, the extreme steering angle $\beta$ increases. In the case of $\omega_n =\pm 100$ rpm, the extreme steering angle is 21 deg. In the case of $\omega_n =\pm 500$ rpm, the extreme steering angle $\beta$ is 35 deg when the angle of attack is 45 deg. In the case of $\omega_n =\pm 1000$ rpm, the extreme steering angle $\beta$ is 42 deg when the angle of attack is 45 deg. The steering angle $\beta$ represents the change in momentum in the YZ-plane, and when the angle of attack is 45 deg, the spacecraft will rotate counterclockwise by 42 deg due to the lift force $F_y$, returning to the state where the angle of attack is 0 deg($\theta$ =180 deg).

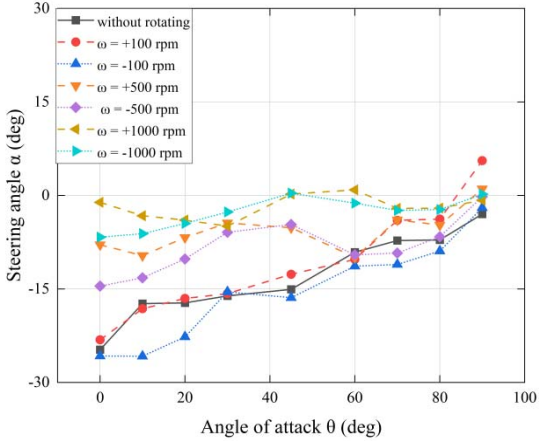

FIGURE 9: Steering angle $\alpha$ of rotating magnetic sail versus the angle of attack $\theta$.

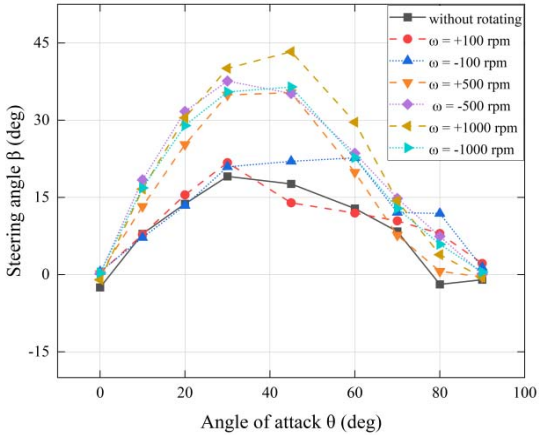

FIGURE 10: Steering angle $\beta$ of rotating magnetic sail versus the angle of attack $\theta$.

*(3) Torque About the X Axis.* The Figure 11 shows the distortion around the X-axis as a function of the angle of attack $\theta$, while the Figure 3 shows the torque around the X-axis defined as the pitching moment applied on the magnetic sail. According to the right-hand (screw) rule defining the direction of torque, clockwise is positive. The X-axis torque applied on the sail is calculated using the conservation of angular momentum of the ions in the simulation domain. First, the change in ion angular momentum with time is calculated. Secondly, the reaction force (X-axis torque) is calculated as **r**×**F**, where **r** is the position vector from the force position to the coil center and **F** is the time variation of the linear momentum calculated within the simulation domain. In the absence of rotation, the maximum torque is obtained at $\theta$ = 45 deg. As shown in the Figure 10, the maximum torque is obtained at $\theta$ = 30 deg for the +$\omega_n$ condition and at $\theta$ = 45 deg for the -$\omega_n$ condition. Note that the torque around the X-axis disappears when the angle of attack is 0 deg or 90 deg, while for other angles of attack, the asymmetry of the magnetic field causes the particle flow to become asymmetric, leading to the torque around the X-axis applied on the coil. In the absence of rotation and for the -$\omega_n$ condition, a clockwise (positive) torque around the X-axis will always be generated, making the magnetic sail return to 0 deg for other angle of attack states (0 deg to 90 deg). However, for the +$\omega_n$ condition, a counterclockwise (negative) torque around the X-axis may be generated at $\theta \geq$ 60 deg, which causes the magnetic sail to return from the current angle of attack to 90 deg. The magnitude of the torque is affected by the rotation speed, and the faster the rotation speed, the greater the torque. Therefore, when the magnetic sail is at $\theta$ = 0 deg, if the angle of attack changes due to external slight disturbance (e.g., changing to 10 deg), it will return to 0 deg due to the clockwise torque around the X-axis. If the magnetic sail is at $\theta$ = 90 deg, if the angle of attack changes due to external slight disturbance (e.g., changing to 80 deg), it will decrease from the disturbance angle to 0 deg under the -$\omega_n$ condition due to the clockwise torque around the X-axis. Conversely, under the +$\omega_n$ condition, it may return from the disturbance angle to 90 deg due to the counterclockwise torque around the X-axis. These results indicate that the stability of a rotating magnetic sail is similar to that of a non-rotating magnetic sail, meaning that $\theta$ = 0 deg or $\theta$ = 180 deg is the most stable angle of attack. Additionally, rotation causes the magnetic sail to increase its rotation angle after being subjected to external disturbance, allowing it to return to a stable state ($\theta$ = 0 deg or $\theta$ = 90 deg) more quickly.

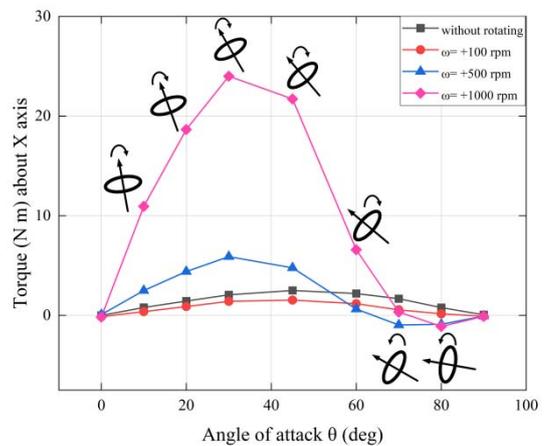

(a) +$\omega_n$

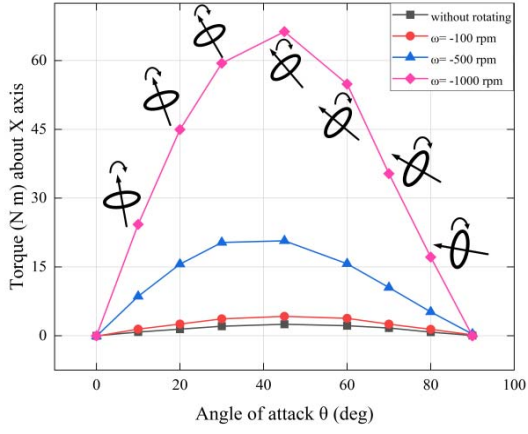

(b)-$\omega_n$

FIGURE 11: X-axis torque exerted on magnetic sail versus the angle of attack $\theta$ for rotational speed $\omega_n$=0, ±100, ±500, ±1000 rpm.

*3.3. Thrust Characteristics Depending on $u_i$.* In this section, the dependence of thrust on the ion incoming velocity $u_i$ is studied. With $M_m = 10^8$ Am$^2$ fixed, ion incoming velocities are set to $5\times10^4$ m/s, $5\times10^5$ m/s, $1\times10^6$ m/s, and $5\times10^6$ m/s respectively, and thrust is considered for an attack angle θ of 0 deg. The results are shown in the Figure 12, where (a) represents the +$\omega_n$ case and (b) represents the -$\omega_n$ case. It can be seen from the figure that in the high-speed range of $u_i$=$5\times10^6$m/s, the thrust is independent of rotation speed, which means that the efficiency of the rotating magnetic sail is less efficient within this velocity range. The +$\omega_n$ case shows good response in the medium-speed range ($u_i$=$5\times10^5$m/s, $1\times10^6$m/s) and can significantly enhance the thrust of the magnetic sail, with the thrust gain proportional to the rotation speed. The -$\omega_n$ case responds to the low-speed range ($u_i$=$5\times10^4$m/s), with the thrust gain proportional to the rotation speed, and about three times the thrust can be obtained under the condition of +$\omega_n$. Therefore, it can be inferred that the thrust gain brought about by rotation decreases with increasing incoming velocity. For higher initial velocities, the decrease in interaction amplitude brought about by rotation can be explained by their smaller available interaction time scales. Since the initial velocity is higher, the rotation speed is not sufficient to cause an increment in the twisted particle motion trajectory, and the interaction time is shorter for high-speed particles under the same conditions, so the interaction will be less. Conversely, for lower speeds, rotation speed may be the primary factor determining the particle's direction of motion, resulting in more complex particle trajectories and a larger interaction window.

After obtaining the approximate value of thrust under stable attack angles, the next step is to examine the correlation between thrust and sail attack angles (0 deg to 90 deg) under high-speed and low-speed incident. The Figure 13 shows a three-dimensional surface plot and projection plot of the thrust dependent on incoming velocity and attack angle. The markers represent numerically fitted three-dimensional surfaces generated through simulations, and the legend color represents the magnitude of the thrust. The form of the force-velocity profile changes with different attack angles.

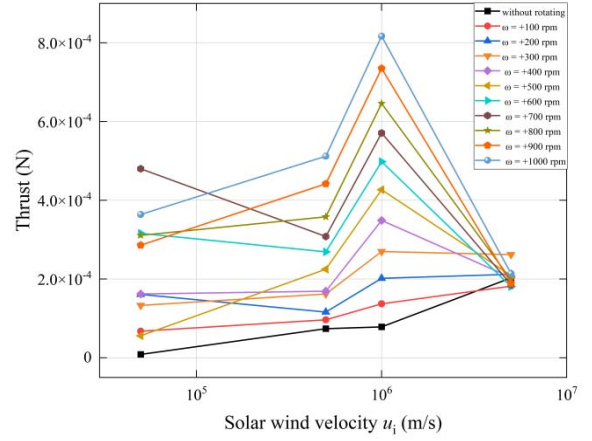

(a) +$\omega_n$

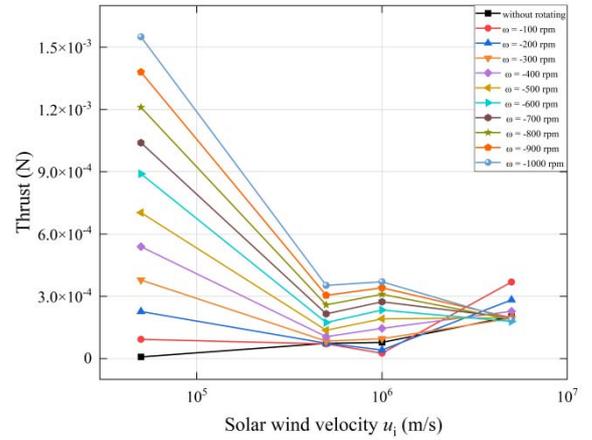

(b)-$\omega_n$

FIGURE 12: Thrust versus the incoming solar wind velocity and rotational speed $\omega_n$.

In the case of no rotation, the θ = 0 deg configuration has no local maximum and the function is monotonically increasing within the investigated range. However, the θ = 90 deg configuration has a local maximum at $u_i$ = 5 × 10$^5$ m/s. By comparing with other angles, it can be seen that as the angle decreases, a larger thrust can be obtained at higher incident speeds. Therefore, it is expected that the thrust of the non-rotating magnetic sail will increase with increasing incident ion velocity.

The situation of rotation is analyzed below. At $\omega_n = +1000$ rpm, the maximum thrust is still obtained with the $\theta = 0$ deg configuration, but the thrust is no longer positively correlated with the incoming speed. The thrust has a local maximum at $u_i = 5\times10^4$ m/s, $1\times10^6$ m/s. In addition, in the low-speed range ($u_i = 5\times10^4$ m/s), the direction of the thrust may change to the -Z direction for $\theta = 60$ deg to $\theta = 90$ deg. This may be because for lower incoming velocity, the electromagnetic force applied to the particles is is very small, while the momentum increment obtained by rotation is larger and opposite in direction to the electromagnetic force, which counteracts the momentum brought by the electromagnetic force.

At $\omega_n = -1000$ rpm, the maximum thrust gain is obtained at $u_i = 5\times10^4$ m/s, and the function is monotonically decreasing within the investigated speed range. It can be seen that as the angle decreases, the maximum thrust can be obtained at lower speeds. This suggests that the magnetic sail with $-\omega_n$ rotation has a better braking effect on solar wind ions in the low-speed range.

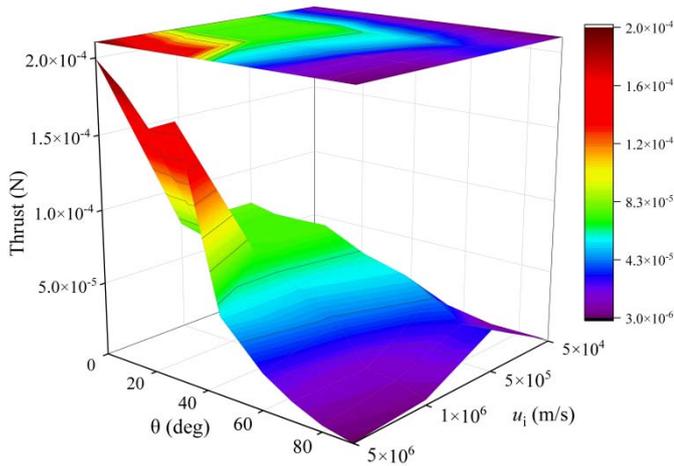

(a) $\omega_n = 0$ rpm

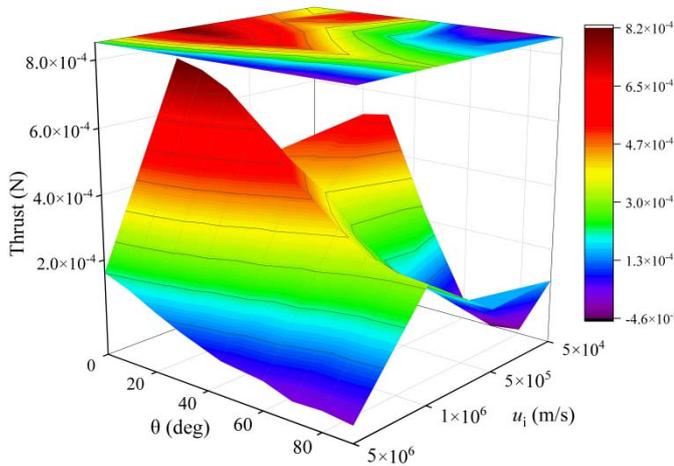

(b) $\omega_n = +1000$ rpm

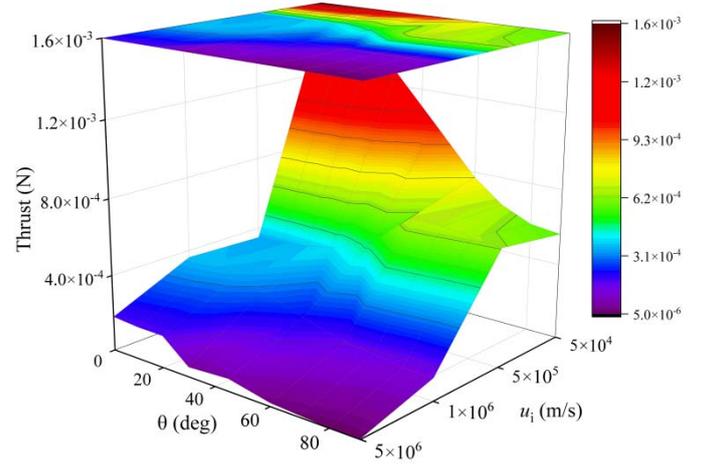

(c) $\omega_n = -1000$ rpm

FIGURE 13: Thrust as a function of the angle of attack and the incoming ion velocity.

*3.4. The Rotating Magnetic Sail in Planetary Orbit.* The application of rotating magnetic sail in earth orbit is analyzed. Table 2 lists the assumed parameters for different orbits simulated by the rotating magnetic sail[21]. Figure 14 shows the thrust levels achieved by magnetic sails of four different sizes in different orbits. First, consider the case of operating in LEO(Low Earth Orbit) at an altitude of 400 km, which is used for propulsion in the orbital direction, assuming that the spacecraft is moving at a speed of 8 km/s[21]. In the simulation, the ion incoming velocity is assumed to be equal to the spacecraft velocity within the control volume surrounding the satellite. In the exosphere region, the typical wind speed of the plasma under quiet geomagnetic conditions ranges from 100 to 300 m/s[22]. Since the wind speed is an order of magnitude smaller than the satellite velocity, the effect of the plasma wind speed is neglected in the calculation[21]. At this altitude, the average ion density is $10^{11}/m^3$. The plasma density is calculated as a function of altitude and time using the International Reference Ionosphere (IRI) 2020 model[23]. In this simulation, the altitude is set to 400 km and daytime conditions are assumed. It is assumed that all ions are O+. This is because at an altitude of 400 km, O+ accounts for over 90% of the plasma, and other ion species contribute very little to the thrust. The temperature of ions and electrons is also obtained from the IRI model. Note that because the mass of O+ is 16 times that of a proton, its number density is much larger than that of solar wind ions (1 AU, $n_i=5\times10^6/m^3$), and the relative velocity of ions in LEO is also smaller. Taking these factors into account, the radius of the rotating magnetic sail coil operating in LEO orbit can be reduced to 1m ($I_{coil}=10^4$ A) to achieve a thrust level comparable to that of a 100m-radius rotating magnetic sail operating in interplanetary space (1 AU). It is worth noting that the ion densities in low orbits around Mars, Venus, and Jupiter are comparable to the above ion density. For example, Mars has a density of $5\times10^{11}/m^3$ at an altitude of 300 km, Venus has a density of $3\times10^{11}/m^3$ at an altitude of 150 km, and Jupiter has a density of $10^{11}/m^3$ at an altitude of 200 km[24], making the use of a rotating magnetic sail around these planets attractive.

Next, we continue to analyze the operation of the rotating magnetic sail in MEO(Middle Earth Orbit) and GEO(Geostationary Orbit). As the orbit is raised to a higher altitude, the ion density drops sharply, and the orbital velocity also decreases. At this altitude, the main component of the plasma in MEO at around 10,000 km is protons, with a density of approximately $10^9/m^3$. In GEO, the ion density is only about $10^7/m^3$, while the orbital velocity is 3 km/s[24]. The combination of these factors requires a larger radius of the coil and current. As shown in Figure 14, a rotating magnetic sail with a radius of 10 meters is needed to achieve millinewton-level thrust. Note that the above orbital analysis does not consider the influence of the Earth's magnetic field on the attitude and torque of the rotating magnetic sail, but only ensures that the magnetic field strength of the artificial dipole at the top of the magnetic layer is greater than the Earth's magnetic field strength at that location. In summary, the rotating magnetic sail has good applicability in Earth orbits with high plasma density and low velocity.

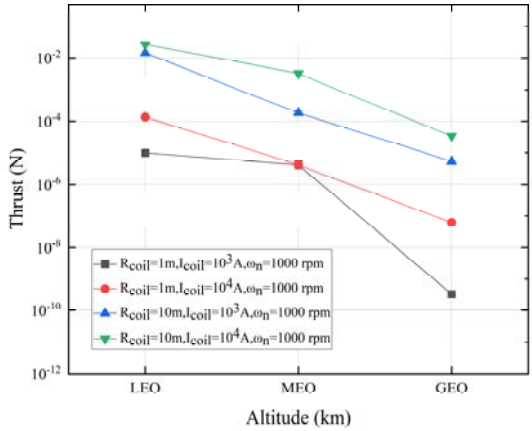

FIGURE 14: Thrust for various altitudes and coil sizes.

TABLE 2: Parameters assumed for the simulation in Planetary Orbit[21].

| | Orbit parameters | | | | Sail parameters | | | |
|---|---|---|---|---|---|---|---|---|
| Orbit | Altitude/km | Ion density/$m^3$ | Ion species | Geomagnetic field/μT | Satellite velocity/m·$s^{-1}$ | Coil radius /m | Coil current/A | $B_{mp}$/μT |
| LEO | 400 | $10^{11}$ | O+ | 50 | 8000 | 1 | 10000 | 314.2 |
| MEO | 10000 | $10^9$ | proton | 20 | 4000 | 10 | 1000 | 31.4 |
| GEO | 36000 | $10^7$ | proton | 7.5 | 3000 | 10 | 1000 | 31.4 |

## 4. Conclusions

In this work, the optimization design of a magnetic sail is proposed, and the concept of a rotating magnetic sail is introduced. The interaction between the solar wind plasma and the magnetic field of the rotating magnetic sail is studied, and 3-D single-component particle numerical simulations are conducted. The purpose of the simulations is to predict the thrust characteristics and thrust-dependent factors of the rotating magnetic sail. The thrust of non-rotating magnetic sail obtained by simulation is compared with that predicted by Full-PIC simulation, and the proportionality of thrust with changes in magnetic moment and size is found to be consistent, verifying the effectiveness of the single-particle model. The research results are summarized as follows:

1) Particle deflection and wake regions are observed around the rotating magnetic sail, and a stable electromagnetic force is generated. The thrust obtained from simulations of the sail at different magnetic moments and rotation speeds verified the trend that the thrust of the rotating magnetic sail increases with increasing rotational speed. The maximum thrust gain ratio obtained at the $\theta=0$ deg configuration is $F_{rotating, \omega=+1000\ rpm}/F_{without\ rorating}=7.88$, while that obtained at the $\theta=90$ deg configuration is $F_{rotating, \omega=-1000\ rpm}/F_{without\ rotating}=19.38$.

2) The dependence of the thrust, steering angle, and X-axis torque of the rotating magnetic sail on the attack angle is explained. The dependence of thrust on the attack angle is related to the initial rotation direction of the sail. When the rotation direction is counterclockwise, the maximum thrust is obtained at an attack angle of $\theta=0$ deg, and vice versa at an attack angle of $\theta=90$ deg. This dependence difference is attributed to the influence of the rotation direction on the vector synthesis of particle velocity in three-dimensional direction, resulting in a change in thrust component. At other attack angle configurations (0 deg<$\theta$<90 deg), due to the asymmetry of the rotating magnetic field, the dependence of turning angle and X-axis torque on sail attack angle is complex. The increase in rotation speed caused the yaw turning angle (atan$F_x/F_z$) of the sail to decrease and the pitch stability (atan$F_y / F_z$) to increase. When the sail is at other attack angles (0 deg<$\theta$<90 deg), it always returns to $\theta$ = 0 deg or $\theta$ = 90 deg through the torque (pitch moment) around the X-axis.

3) The rotating magnetic sail has a good application prospect on orbits with low ion velocity and high number density. Considering the high ion density and low velocity on Earth's orbit, the coil size of the rotating magnetic sail can be greatly reduced. Taking a supply current of 1000A as an example, a sail with a radius of 10m is required to obtain a thrust of $10^{-4}$ N at the height of the GEO, while the radius can be reduced to 1m at the height of the LEO.

In order to fully describe the efficiency of the rotating magnetic sail, further research is needed, including consideration of the effects of electron pressure calculation. The PIC encoding is planned to be used to quantify the effects of electrons and induced electric fields on the thrust vector. In addition, although the rotating magnetic sail has significantly improved the thrust-to-weight ratio compared to the original magnetic sail, further optimization of the design of the rotating magnetic sail is necessary for its actual application in interplanetary orbits. For example, considering the design of a multi-coil dipole magnetic field and the array form of the coil to reduce the size of the rotating magnetic sail.

## Data Availability

The data used to support the findings of this study are available from the corresponding author upon request.

## Conflicts of Interest

All authors declare no possible conflicts of interests.

**Funding:** This work was supported the National Natural Science Foundation of China (42241148, and 51877111).

## Authors' Contributions

M.W. Xu assisted the simulation and wrote the paper. R.H. Quan proposed the design, carried out the simulation and is the corresponding author. Y.J. Yao participated in the writing.

## Acknowledgments

We thank to G.P. Zhu for help in the simulation.